\documentclass[11pt,fleqn]{article}
\usepackage{epsfig,graphics,color}
\definecolor{Black}{named}{Black}
\definecolor{Blue}{named}{Blue}
\definecolor{Red}{named}{Red}
\definecolor{Green}{named}{ForestGreen}
\definecolor{Black}{named}{Black}
\definecolor{Olive}{named}{OliveGreen}
\definecolor{Royal}{named}{RoyalBlue}
\definecolor{Orange}{named}{YellowOrange}
\definecolor{Yellow}{named}{Goldenrod}
\definecolor{Cornblue}{named}{CornflowerBlue}
\definecolor{Lila}{named}{DarkOrchid}

\begin{document}

\null
\hfill IPPP-03-12\\
\null
\hfill DCPT-03-24\\
\null
\hfill hep-ph/0303234

\vskip .4cm

\begin{center}
{\Large A Future Linear Collider with Polarised Beams\\
Searches for New Physics\footnote{Invited plenary talk given at
15th International Spin Physics Symposium,
Sept. 2002,\\\phantom{LH.}  BNL, Brookhaven, USA.}
}

\vskip .6cm
{\large Gudrid Moortgat-Pick\footnote{email: g.a.moortgat-pick@durham.ac.uk.}}

\vskip .3cm
{\footnotesize\it 
\noindent
IPPP, University of Durham, Durham DH1 3LE, U.K.}\footnote{Address before 
January 2003: 
DESY, Deutsches Elektronen--Synchrotron, D--22603\\
 \phantom{LH.} Hamburg, Germany.}

\begin{abstract}
There exists a world--wide consensus for
a future $e^+e^-$ Linear Collider in the energy range between
$\sqrt{s}=500-1000$~GeV as the next 
large facility in HEP. The Linear Collider 
has a large physics potential for the
discovery of new physics beyond the Standard Model
and for precision studies of the Standard Model itself.
It is well suited to complement and extend 
the physics program of the LHC.
The use of polarised beams at a Linear Collider will be one of
the powerful tools. In this paper some highlights
of searches for physics beyond the Standard Model 
at a future Linear Collider with polarised $e^-$ and $e^+$ beams
are summarised.
\end{abstract}
\end{center}

\section{Beam Polarisation at a Linear Collider}
\begin{sloppypar}
The next future large experiment in high energy physics will most probably
be a future Linear Collider (LC) in the energy range between LEP and
{\cal O}(1~TeV).  The existing world--wide proposals are designed with high
luminosity of about ${\cal L}=3.4\cdot 10^{34}$cm$^{-2}$s$^{-1}$ at
$\sqrt{s}=500$~GeV and ${\cal L}=5.8 \cdot 10^{34}$cm$^{-2}$s$^{-1}$
at $\sqrt{s}=800$~GeV, see e.g. \cite{TDR}.  A LC will not only be
well suited to
complement \cite{LHC/LC} but also to extend the physics program of the 
Hadron Colliders, the Tevatron and the future Large Hadron Collider 
(LHC).
\end{sloppypar}

A LC has a large potential for the discovery of new particles  
and is -- due to its clear signatures -- very well suited for the 
precise analysis of
new physics (NP) as well as of the Standard Model (SM)
\cite{TDR}.
Providing precision studies in this energy range, in
particular in the electro--weak sector, even small traces of 
physics beyond the SM might be found, even if high scale particles of a NP
might not be directly produced at the LHC or the first phase of a LC. 
For these studies the GigaZ 
option of the LC, i.e. running with very high luminosity 
at the $Z$ and the $WW$ threshold, is decisive.

An important tool of a LC
is the use of polarised beams. In the following we will summarise 
some highlights of searches and analyses of new physics with the help of 
polarised beams \cite{TDR,Steiner,orange}. 

Already in the base line design it is foreseen to use electron
beams polarised to around $80\%$ via a strained photocathode
technology \cite{TDR} similar to those at the SLC where in the
last year of running 1994/95 $P_{e^-}=(77.34\pm 0.61)\%$ (\cite{Steiner}
and references therein) was reached. 
In order to generate also polarised positrons 
the use of a helical undulator is favoured producing 
polarised photons which generate via pair production positrons
with a designed polarisation degree
of about 40\% (with full intensity of the $e^+$ beam) up to 60\% 
(with probably about 55\% intensity) \cite{Floettmann}. 
There already exists a world--wide collaboration supporting the 
activities to get a prototype for a polarised positron source
at the 50 GeV Final Focus Test Beam at SLAC
\cite{Project}. 

It is foreseen to measure the polarisation with Compton polarimetry
and it is assumed that one could reach even an accuracy better than
$\Delta(P_{e^{\pm}})< 0.5\%$ \cite{Schuler}.  The simultaneous use of
M$\not\!o$ller polarimetry will also be studied at the LC
\cite{Gideon}.  However, the reachable accuracy with Compton and
M$\not\!o$ller polarimetry will not be 
sufficient for the high precision
tests at GigaZ. For this purpose one uses an alternative Blondel
Scheme \cite{Blondel}, 
see next section, where one expresses the polarisation via
polarised cross sections. Therefore one can avoid 
absolute measurements of polarisation and uses polarimetry 
only for relative measurements.

After a short introduction into the physics of beam polarisation we
begin our summary with high precision studies of the SM as a
motivation for new physics searches.

\subsection{Introductory remarks}
Within the Standard Model (SM) only $(V-A)$ couplings happen in the 
s--channel and therefore the configurations $LR$ and $RL$ are possible
for the $e^-e^+$ helicities. That means that once the $e^-$ polarisation
is chosen also the $e^+$ polarisation is fixed. For these processes
an additional simultaneous positron polarisation leads to an enhancement
(or suppression) of the
fraction of the colliding particles, which is expressed by
the effective luminosity
\begin{equation}
{\cal L}_{eff}/{\cal L}:=\frac{1}{2}(1-P_{e^-}P_{e^+})
\label{lum_eff}
\end{equation} 
and of the effective polarisation
\begin{equation}
P_{eff}:=(P_{e^-}-P_{e^+})/(1-P_{e^-}P_{e^+}).
\label{pol_eff}
\end{equation}
In Table~\ref{tab_desch} we list ${\cal L}_{eff}$ and $P_{eff}$ 
for some characteristic values of $P_{e^-}$ and $P_{e^+}$.
One can see that even with a
completely polarised $e^-$ beam the fraction of colliding
particles is not enhanced, however, with simultaneously
polarised positrons this fraction will be enhanced.

It is well--known that with suitably polarised beams one can 
suppress background processes, e.g.
the dominating
SM backgrounds $e^+e^-\to W^+W^-$ and $ZZ$. 
Some scaling factors $\sigma^{pol}/\sigma^{unpol}$ for these
processes are given in Table~\ref{tab_back}.

Moreover, beyond the SM there are also coupling structures in the s-channel
possible where also the configurations $LL$ and $RR$ could lead to 
strong signals. Simultaneous polarisation of both beams would lead therefore
to fast and easy diagnostics and we will give one example.

\begin{table}
\begin{center}
\begin{tabular}{|l||c|c|c|c||c|c|}
\hline
& RL & LR & RR & LL & $P_{eff}$ & 
${\cal L}_{eff}/{\cal L}$ \\ \hline
$P_{e^-}=0$, & 0.25 & 0.25 & 0.25 & 0.25 & 0. & 0.5\\ 
$P_{e^+}=0$ &&&&&& \\ \hline
$P_{e^-}=-1$, & 0 & 0.5 & 0 & 0.5 & $-1$ & 0.5 \\
$P_{e^+}=0$ &&&&&&\\ \hline 
$P_{e^-}=-0.8$, & 0.05 & 0.45 & 0.05 & 0.45 & $-0.8$ & 0.5 \\
$P_{e^+}=0$ &&&&&&\\ \hline 
$P_{e^-}=-0.8$, & 0.02 & 0.72 & 0.08 & 0.18 & 
$-0.95$ & 0.74 \\
$P_{e^+}=+0.6$ &&&&&& \\ \hline 
\end{tabular}
\caption{ Fraction of colliding particles (${\cal L}_{eff}/{\cal L}$)
and the effective polarisation ($P_{eff}$) 
for different beam polarisation configurations, which are characteristic
for (V-A) processes in the s--channel \cite{talk_desch}.
\label{tab_desch}}
\end{center}
\end{table}
\begin{table}
\begin{center}
\begin{tabular}{|c|c|c|}\hline
($P_{e^-}=\mp 80\%$, $P_{e^+}=0,\pm 60\%$)
& $e^+ e^-\to W^+ W^-$ & $e^+ e^-\to ZZ$\\[.2em] \hline
$(+0)$ & 0.2 & 0.76 \\
$(-0)$ & 1.8 & 1.25 \\ \hline
$(+-)$ & 0.1 & 1.05 \\
$(-+)$ & 2.85 & 1.91\\ \hline
\end{tabular}
\caption{Scaling factors $\sigma^{pol}/\sigma^{unpol}$ for the
dominating SM background processes $e^+e^-\to W^+W^-$ and $ZZ$ for
different configurations of beam polarisation \label{tab_back} \cite{Steiner}.}
\end{center}
\end{table}
These given 'rules' are not generally valid for t--channel exchanges
since in that case the helicity of the incoming $e^-$ is only coupled to the
outgoing particle at the vertex and not to the incoming $e^+$.  This
can be easily seen when studying the well--known Bhabha background.
For small energies the s--channel with its $LR$ and $RL$ coupling
characteristics dominates. However, for higher energies, also $LL$ 
coupling is possible via the $\nu$--exchange in the t--channel.
Another example, where the use of both beams polarised is obvious, 
is the single $W$ background since with $e^-$ polarisation only
the $W^-$ signal can be suppressed. For the corresponding signal from
$W^+$ the polarisation of $e^+$ is needed.

\section{Electroweak high precision analyses of the SM}
Electroweak precision tests with an unprecedented accuracy -- at
high energies as well as at the $Z$ resonance 
and the $WW$ threshold -- would allow to see
hints for new physics, even if new particles are not directly produced. 
In the following section 
we list some examples for these high precision measurements at a LC, e.g.
the measurement of triple gauge couplings. After that we
will have a look at the additional prospects of GigaZ which is planned as an
upgrade.

\subsection{Anomalous couplings in $e^+e^-\to~W^+W^-$}
In order to test the SM with high precision one can 
carefully study triple gauge boson couplings, which are generally
parametrised in an effective Lagrangian e.g. 
by the C--, P--conserving couplings
$g_1^V$, $\kappa_V$, $\lambda_V$ with 
$V=\gamma, Z$.  In the SM at tree level the couplings
have to be $g_1^V=1=\kappa_V$, while $\lambda_V$ are identical to zero.

These couplings can be determined by measuring the angular
distribution and polarisation of the $W^{\pm}$'s.
Simultaneously fitting of all couplings results in a strong 
correlation between the $\gamma-$ and $Z-$couplings. It turns out
that the polarisation of the beams is very powerful for separating these 
couplings:
e.g. the polarisation of $P_{e^-}=\pm 80\%$ (together with
$P_{e^+}=\mp 60\%$) improves the sensitivity up to a factor 1.8 (2.5),
see Table~\ref{tab_menges} \cite{Moenig,TDR}.

\begin{table}
\begin{center}
\begin{tabular}{|crrrrr|}
\hline
error [$10^{-4}$]:& 
 $\Delta g^1_Z$ & $\Delta\kappa_{\gamma}$ & $\lambda_{\gamma}$  & 
$\Delta\kappa_Z$ & $\lambda_Z$  \\
\hline\hline
\multicolumn{6}{|c|}{unpolarised beams} \\ \hline
$\sqrt{s}=500$~GeV & 38.1 & 4.8 & 12.1 & 8.7 & 11.5 \\
$\sqrt{s}=800$~GeV & 39.0 & 2.6 &  5.2 & 4.9 &  5.1  \\
\hline
\multicolumn{6}{|c|}{only electron beam polarised, $|P_{e^-}|=80\%$}  
\\ \hline
$\sqrt{s}=500$~GeV & 24.8 & 4.1 &  8.2 & 5.0 &  8.9 \\
$\sqrt{s}=800$~GeV & 21.9 & 2.2 &  5.0 & 2.9 &  4.7 \\
\hline
\multicolumn{6}{|c|}{both beams polarised, $|P_{e^-}|=80\%$, 
$|P_{e^+}|=60\%$} \\ \hline
$\sqrt{s}=500$~GeV & 15.5 & 3.3 &  5.9 & 3.2 &  6.7  \\
$\sqrt{s}=800$~GeV & 12.6 & 1.9 &  3.3 & 1.9 &  3.0  \\
\hline
\end{tabular}
\caption{Sensitivity for anomalous triple gauge couplings with different 
configurations of beam polarisation \cite{TDR}.\label{tab_menges}}
\end{center}
\end{table}

\subsection{Transversely polarised beams in $e^+e^-\to W^+ W^-$}
Another promising possibility to study the origin of
electroweak symmetry breaking
is the use of transversely polarised $e^+e^-$ beams which projects out
$W^+_L W^-_L$ \cite{Karol}.  The asymmetry with respect
to the azimuthal angle of this process focusses on the $LL$ mode.
This asymmetry is very pronounced at high energies reaching about $10\%$.
The advantage of this observable is that at high energies this asymmetry
peaks at larger angles and not
in beam direction where the analysis might be difficult.  One has
to note, however, that for the use of transverse
beams the polarisation of 
both beams is needed. The effect does not occur if only one 
beam is polarised since the cross section is given by:
\begin{eqnarray}
\sigma&=&(1-P_{e^-}^L P_{e^+}^L)\sigma_{unp}+(P_{e^-}^L-P_{e^+}^L)
\sigma_{pol}^L+P_{e^-}^T P_{e^+}^T \sigma_{pol}^T.
\label{eq_trans}
\end{eqnarray}

\subsection{GigaZ}
\begin{figure}[htb]
\begin{picture}(10,15)
\setlength{\unitlength}{1cm}
\put(-.2,-6.5){\mbox{\includegraphics[height=.33\textheight]{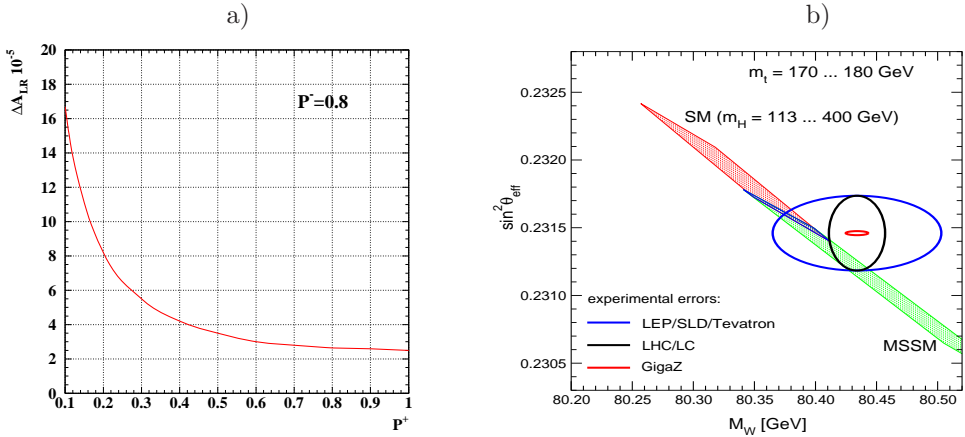}}}
\put(2.8,-1){\small a)}
\put(10.5,-1){\small b)}
\put(6.2,-6.5){\mbox{
\includegraphics[height=.27\textheight,width=.33\textheight]{georg}}}
\end{picture}\par\vspace{6.3cm}
\caption{ a) Test of Electroweak Theory: The statistical error on the
left--right asymmetry $A_{LR}$ of $e^+e^-\to Z\to\ell \bar{\ell}$ at
GigaZ as a function of the positron polarisation $P_{e^+}$ for fixed
electron polarisation $P_{e^-}=\pm 80\%$ \cite{Moenig}. In b) The
theoretical prediction for the relation between $\sin^2\theta{eff}$
and $M_W$ in the SM for Higgs boson masses in the intermediate range
is compared to the experimental accuracies at LEP 2/Tevatron (Run
IIA), LHC/LC and GigaZ \cite{gigaz}.
\label{fig_gigaz}}
\end{figure}

At the GigaZ option 
$e^+e^-\to Z\to f \bar{f}$ is studied and the
effective 
electroweak leptonic mixing angle can be measured via the left--right
asymmetry 
\begin{equation}
A_{LR}=\frac{2(1-4 \sin^2\Theta^{\ell}_{eff})}
{1+(1-4 \sin^2\Theta^{\ell}_{eff})^2}
\label{eq_alr}
\end{equation}
of this process. 
Since one gets only a gain in 
statistical power if the error due to the polarisation
measurement $\Delta A_{LR}(pol)$
is smaller than the statistical error
$\Delta A_{LR}(stat)$ one has to know $P_{e^{\pm}}$ extremely accurately.
Up to now even $\Delta P_{e^{\pm}}<0.5\%$ would not be sufficient. Therefore
one uses an alternative Blondel Scheme \cite{Moenig,TDR} and
expresses $A_{LR}$ via polarised rates:
\begin{equation}
A_{LR}=\sqrt{\frac{(\sigma^{++}+\sigma^{+-}-\sigma^{-+}-
\sigma^{--})(-\sigma^{++}+\sigma^{+-}-\sigma^{-+}+\sigma^{--})}
{(\sigma^{++}+\sigma^{+-}+\sigma^{-+}+\sigma^{--})(-\sigma^{++}+\sigma^{+-}
+\sigma^{-+}-\sigma^{--})}}
\label{eq_blon}
\end{equation}
With this method, polarimetry has to be used only for calibration
and one can reach a spectacular accuracy for the electroweak
observables, see Table~\ref{tab_blond} \cite{TDR}.  The polarisation of
the positron beam is absolutely needed but already a polarisation of
about $P_{e^+}=|40\%|$ would be sufficient, see Fig.~\ref{fig_gigaz}a,
to measure these observables with an unprecedented accuracy.

As an example of the potential of the GigaZ $sin^2\theta_{eff}$ 
measurement, Fig.~\ref{fig_gigaz}b \cite{gigaz}
compares the present experimental accuracy on $sin^2\theta_{eff}$ and $M_W$
from LEP/SLD/Tevatron and the prospective
accuracy from the LHC and from a LC without GigaZ option
with the predictions of the SM and the MSSM. With GigaZ a very sensitive
test of the theory will be possible.

\vspace{-.8cm}

\begin{table}
\begin{center}
\begin{tabular}{c|cccc}
& LEP2/Tev.& Tev./LHC & LC & GigaZ/WW\\\hline
$M_W$ & 34 MeV & 15 MeV & 15 MeV & 6 MeV\\
$\sin^2\theta_{eff}$ & 0.00017 & 0.00017 & 0.00017 & 0.00001\\
$m_t$ & 5 GeV & 2 GeV &0.2 GeV & 0.2 GeV \\
$m_h$ & -- & 0.2 GeV & 0.05 GeV & 0.05 GeV\\[1em]
\end{tabular}
\caption{Sensitivity to electroweak observables at different colliders
in comparison \cite{gigaz}. \label{tab_blond}}
\end{center}
\end{table}

\noindent It should also be mentioned that
with the help of polarised beams a LC could be
sensitive to electroweak dipole form factors.
They have been analysed in
\cite{Stahl} with regard to CP violation of the $\tau$ lepton and via 
CP--odd triple product correlations. 
Sensitivity bounds for the real and imaginary
parts of these form factors have been set 
up to $O(10^{-19})$~ecm.

In the same context one should not forget that also for searches of
heavy gauge bosons, as e.g. for the $Z'$, the use of polarised beams
enhances the discovery range. Also for contact interactions the
sensitivity can be enhanced significantly \cite{Steiner,Riemann}.

\section{Beyond the Standard Model}
\subsection{Supersymmetry}
Supersymmetry is widely regarded as one of the best motivated
extensions of the SM.  However, since the SM particles and their SUSY
partners are not mass degenerate, SUSY has to be broken, which leads
even for its minimal version, the Minimal Supersymmetric Standard
Model (MSSM), to about 105 free new parameters. In specific scenarios of
SUSY breaking one can end with only a few 
parameters: 5 in mSUGRA, 4 in AMSB and
5 in GMSB.  However, one should note that 
one of the most favoured motivation
for SUSY -- the unification of the gauge couplings -- is consistent
within the general MSSM independently of the large number of new parameters.

In order to exactly pin down the structure of the underlying model it
is therefore 
unavoidable to extract the parameters without assuming a particular
breaking scheme. 
Since the LC with its extremely clear signatures provides a measurement 
of the particle masses 
up to $O(100)$~MeV, of the rates and branching ratios at the 
$\%$ level, 
the LC is well suited for revealing the underlying structure of the model.
Different step--by--step procedures have been worked out to
determine the general MSSM parameters and to test
fundamental SUSY assumptions as e.g. the equality of quantum numbers
or of couplings of the particles and their SUSY partners as model 
independent as possible.  It turns
out that the use of polarised beams plays a decisive role in this
context.
\subsubsection{Stop mixing angle in $e^+e^-\to \tilde{t}_1\tilde{t}_1$}
As demonstrated 
in \cite{TDR} the mass and the mixing angle of $\tilde{t}$ can be extracted 
with high precision via the study of polarised cross sections
for light stop production. At a high luminosity LC and with 
$P(e^-)=80\%$ and 
$P(e^+)=60\%$ an accuracy of $\delta(m_{\tilde{t}_1})\approx 0.8$~GeV and
$\delta(\cos\theta_{\tilde{t}})\approx 0.008$ could be reachable, see 
Fig.~\ref{fig_sel}, \cite{Wien}.
Similar studies have been done for the $\tilde{\tau}$ sector \cite{Stau}. 
\subsubsection{Quantum numbers in 
$e^+e^-\to~\tilde{e}^{+}_{L,R}\tilde{e}^-_{L,R}$} 
SUSY transformations associate chiral leptons to their scalar SUSY partners:
$e^-_{L,R}\leftrightarrow \tilde{e}^-_{L,R}$ 
and the antiparticles 
$e^+_{L,R}\leftrightarrow \tilde{e}^+_{R,L}$. In order to prove
this association between scalar particles and chiral quantum numbers the use
of polarised beams is necessary \cite{Bloechi}. The process occurs via
$\gamma$ and $Z$ exchange in the s--channel and via $\tilde{\chi}^0_i$
exchange in the t--channel. As already mentioned in the general 
introduction one has direct coupling between
the SM particle and its scalar partner
only in the t--channel. Therefore one has to project out the t--channel 
exchange in order to test the association of chiral 
quantum numbers to the scalar SUSY partners.

\begin{figure}[htb]
\begin{picture}(15,7)
\setlength{\unitlength}{1cm}
\put(-.8,-5.9){\mbox{\includegraphics[height=.33\textheight]{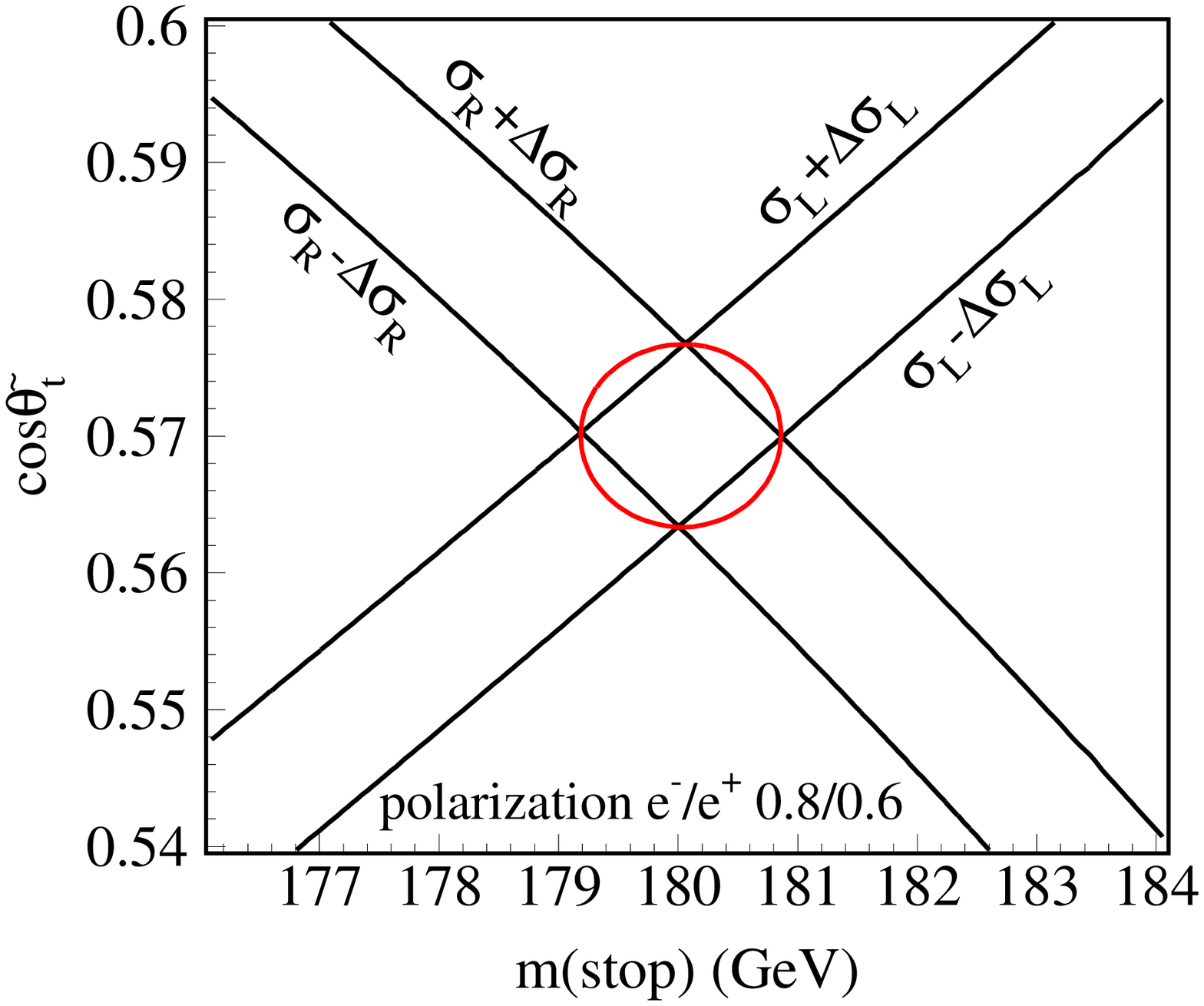}}}
\put(6,-5.4){\mbox{\includegraphics[height=.25\textheight]{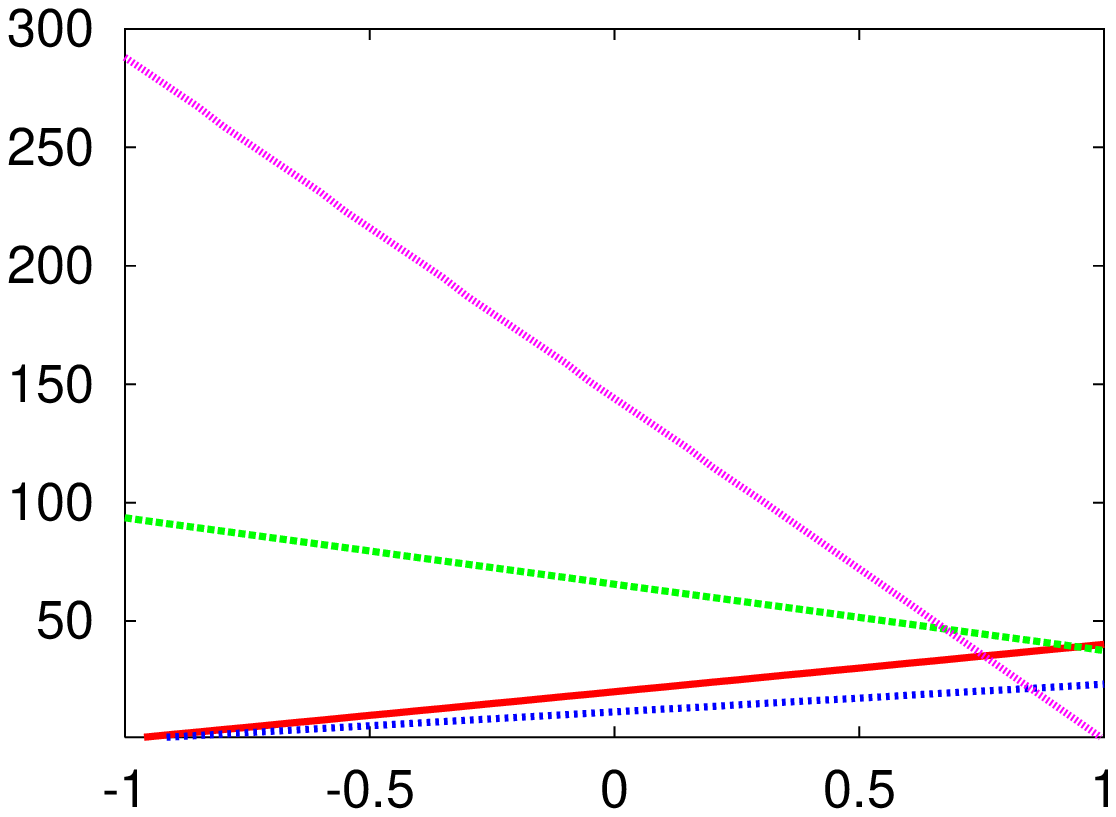}}}
\put(6.4,-.8){\makebox(0,0)[bl]{{$\sigma$[fb]}}}
\put(8.8,-2.9){\makebox(0,0)[br]{{\color{Lila}
$\tilde{e}^-_{L} \tilde{e}^+_{R}$}}}
\put(8.5,-3.6){\makebox(0,0)[br]{{\color{Green}
$\tilde{e}^-_{R} \tilde{e}^+_{R}$}}}
\put(8.3,-4.6){\makebox(0,0)[br]{{\color{Red}
$\tilde{e}^-_{R} \tilde{e}^+_{L}$}}}
\put(13.4,-4.6){\makebox(0,0)[br]{{\color{Blue}
$\tilde{e}^-_{L} \tilde{e}^+_{L}$}}}
\put(12.5,-5.6){\makebox(0,0)[br]{{$P_{e^+}$}}}
\put(12.2,-2.1){\makebox(0,0)[br]{{$P_{e^-}=-80\%$}}}
\put(9.7,-1.5){$\sqrt{s}=400$~GeV}
\put(2.7,-.5){\small a)}
\put(9.5,-.5){\small b)}
\end{picture}\vspace{5.5cm}
\caption{Test of selectron quantum numbers in
$e^+e^-\to\tilde{e}_{L,R}^+\tilde{e}_{L,R}^-$ with fixed electron
polarisation $P(e^-)=-80\%$ and variable positron polarisation
$P(e^+)$. For  $P(e^-)=-80\%$ and $P(e^+)<0$ both pairs
$\tilde{e}_L^- \tilde{e}_R^+$ and
$\tilde{e}^-_R \tilde{e}^+_R$ still contribute. For $P(e^+)=-60\%$
the pair $\tilde{e}_L^- \tilde{e}_R^+$ dominates by more than a factor 3
\cite{TDR,Steiner,Bloechi}.}
\label{fig_sel}
\vspace{-.3cm}
\end{figure}

With e.g. completely polarised 
$e^-_L e^+_L$ only the pair $\tilde{e}^-_L \tilde{e}^+_R$ contributes. 
Due to
their $L,R$ coupling character $\tilde{e}_L$, $\tilde{e}_R$ can be
discriminated via their decay characteristics 
and can be identified via their charge. One
has to note that a polarised $e^+$ beam is necessary. Even completely
polarised $e^-$ would not be sufficient, 
since the s-channel exchange could
not be switched off completely.  However, even 
if only partially polarised beams
of maximal $P_{e^-}=-80\%$ and $P_{e^+}=-60\%$ were available it
could
be sufficient to probe this association, since in this case the pair
$\tilde{e}^-_L\tilde{e}^+_R$ dominates by a factor
of 3 in our example, Fig.~\ref{fig_sel}b, \cite{Bloechi}.

\subsubsection{Gaugino/higgsino sector}
\begin{figure}[htb]
\setlength{\unitlength}{1cm}
\begin{picture}(15,8)
\put(-.5,0){\mbox{
\includegraphics[height=.27\textheight,width=.3\textheight]{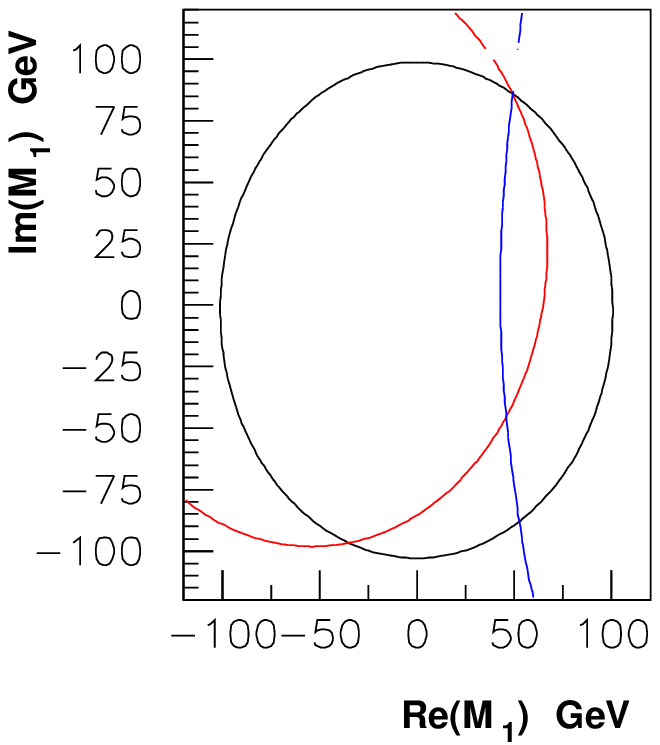}}}
\put(4.5,4.5){\small $m_{\tilde{\chi}^0_1}$}
\put(2.8,2){\small\color{Red} $m_{\tilde{\chi}^0_2}$}
\put(2.7,3.5){\small\color{Blue} $\sigma(\tilde{\chi}^0_1 \tilde{\chi}^0_2)$}
\put(6.1,0){\mbox{\includegraphics[height=.3\textheight,
width=.35\textheight]{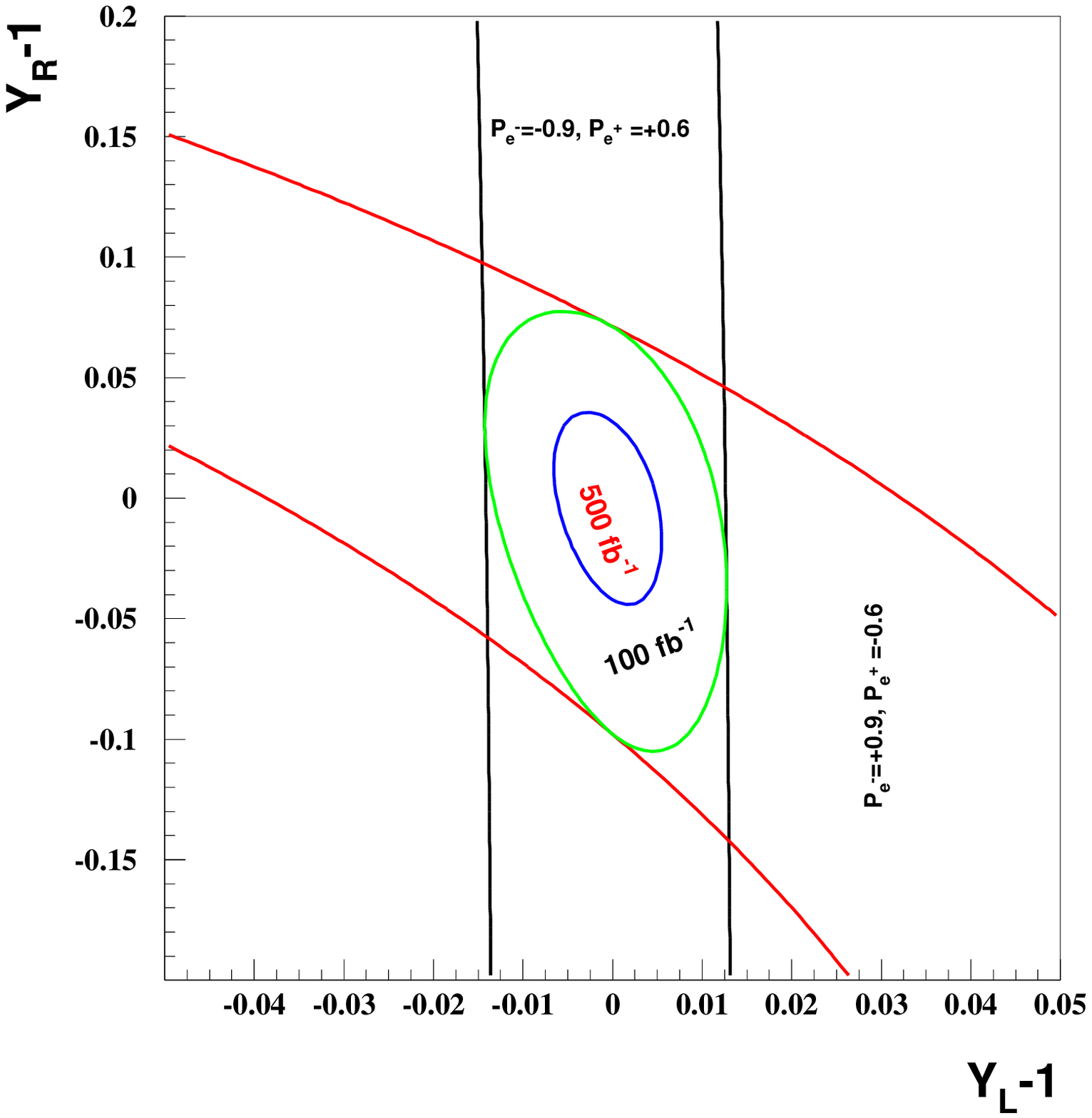}}}
\put(3,5.3){\small a)}
\put(9.5,5.3){\small b)}
\end{picture}
\caption{\label{fig_m1}
The contours of
two neutralino masses (1,2) and one neutralino production cross section
$\sigma(e^+e^-\to\tilde{\chi}^0_1\tilde{\chi}^0_2)$ in the
$Re(M_1)$, $Im(M_1)$ plane;
b) Contours of the cross sections $\sigma(e^+_Re^-_L\to
\tilde{\chi}^0_1\tilde{\chi}^0_2)$ and
       $\sigma(e^+_Le^-_R\to\tilde{\chi}^0_1\tilde{\chi}^0_2)$
in the plane of the Yukawa couplings $g_{\tilde{W}}$
       and $g_{\tilde{B}}$ normalised to the SU(2) and U(1) gauge
       couplings $g$ and $g'$
$\{Y_L=g_{{\tilde W}}/g,\, Y_R=g_{{\tilde B}}/g'\,\}$ \cite{ckmz02}. }
\end{figure}

The SUSY partners of the charged and neutral gauge bosons are the
charginos $\tilde{\chi}^{\pm}_{1,2}$ and neutralinos
$\tilde{\chi}^0_{1,\ldots,4}$.  Since SUSY is a broken the
electroweak eigenstates mix and strategies have been worked out to
determine the mixing angles via polarised rates in $e^+
e^-\to \tilde{\chi}^{\pm}_i\tilde{\chi}^{\mp}_j$ and
$e^+e^-\to\tilde{\chi}^0_i\tilde{\chi}^0_j$ and to derive the underlying
MSSM parameters (\cite{ckmz02} and references therein).  Even if only
the lightest particles
$\tilde{\chi}^+_1\tilde{\chi}^-_1$, $\tilde{\chi}^0_1\tilde{\chi}^0_2$
were accessible, it would be sufficient for determining the
fundamental MSSM parameters $M_1$, $\Phi_{M_1}$, $M_2$ and $\mu$,
$\Phi_{\mu}$, i.e.  the U(1), the SU(2), and the higgsino mass parameters
with its CP--violating phases. The ratio of the two Higgs vev's
$\tan\beta=v_2/v_1$ can only be derived via this sector
if $\tan\beta$ is not too large  (\cite{ckmz02} and references therein).
In Fig.~\ref{fig_m1}a, e.g., it is demonstrated, how to fix $|M_1|$
and $\Phi_{M_1}$ with polarised cross sections
$\sigma(e^+e^-\to\tilde{\chi}^0_1\tilde{\chi}^0_2)$ and the light
masses $m_{\tilde{\chi}^0_{1,2}}$. In this context the beam
polarisation is needed in order to resolve ambiguities and to improve
the statistics.
Once the parameters are determined one can efficiently test 
whether the gauge couplings $g_{Bee}$ and $g_{Wee}$
are identical to the Yukawa couplings
$g_{\tilde{B}e\tilde{e}}$ and $g_{\tilde{W} e \tilde{e}}$, respectively, by
studying the polarised cross sections with a variable ratio of
$g_{Bee}/g_{\tilde{B}e\tilde{e}}$ and $g_{Wee}/g_{\tilde{W} e \tilde{e}}$
and comparing it with experimental values \cite{ckmz02}, see 
Fig.~\ref{fig_m1}b.

\subsubsection{The case of high $\tan\beta$: $\tau$ polarisation}
A crucial parameter is $\tan\beta$, but if $\tan\beta$ is high it will
be very difficult to determine it very accurately.  In case of high
$\tan\beta>10$ the chargino and neutralino sector is insensitive to
this parameter. But even in the Higgs sector the case $\tan\beta>10$
will lead to large uncertainties $\Delta(\tan\beta)>10\%$
\cite{higgs_det}.  However, in many scenarios one could then determine
$\tan\beta$ from another sector whose particles are relatively light:
the $\tilde{\tau}$ sector \cite{Stau}.

\begin{figure}[htb]
\setlength{\unitlength}{1cm}
\begin{picture}(10,5)
\put(-.5,-5){\mbox{
\includegraphics[height=.38\textheight,width=.33\textheight]{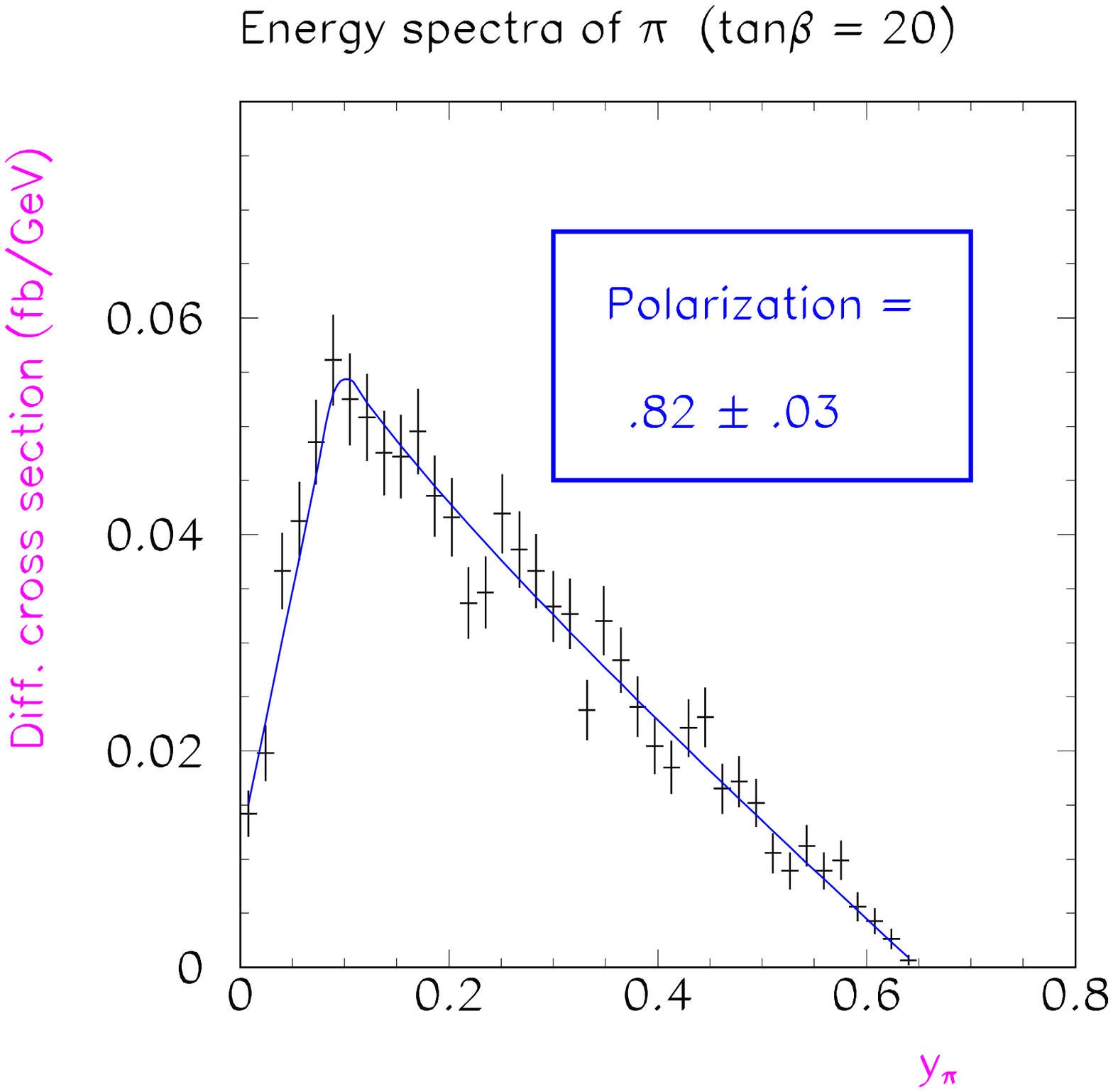}}}
\put(0,2.1){\small a)}
\put(9.3,2.1){\small b)}
\put(6.3,2.1){\small $\tan\beta$}
\put(11.5,-3.1){\small $P_{\tilde{\tau}_1\to\tau}$}
\put(6,-2.8){\mbox{\includegraphics[height=.25\textheight]{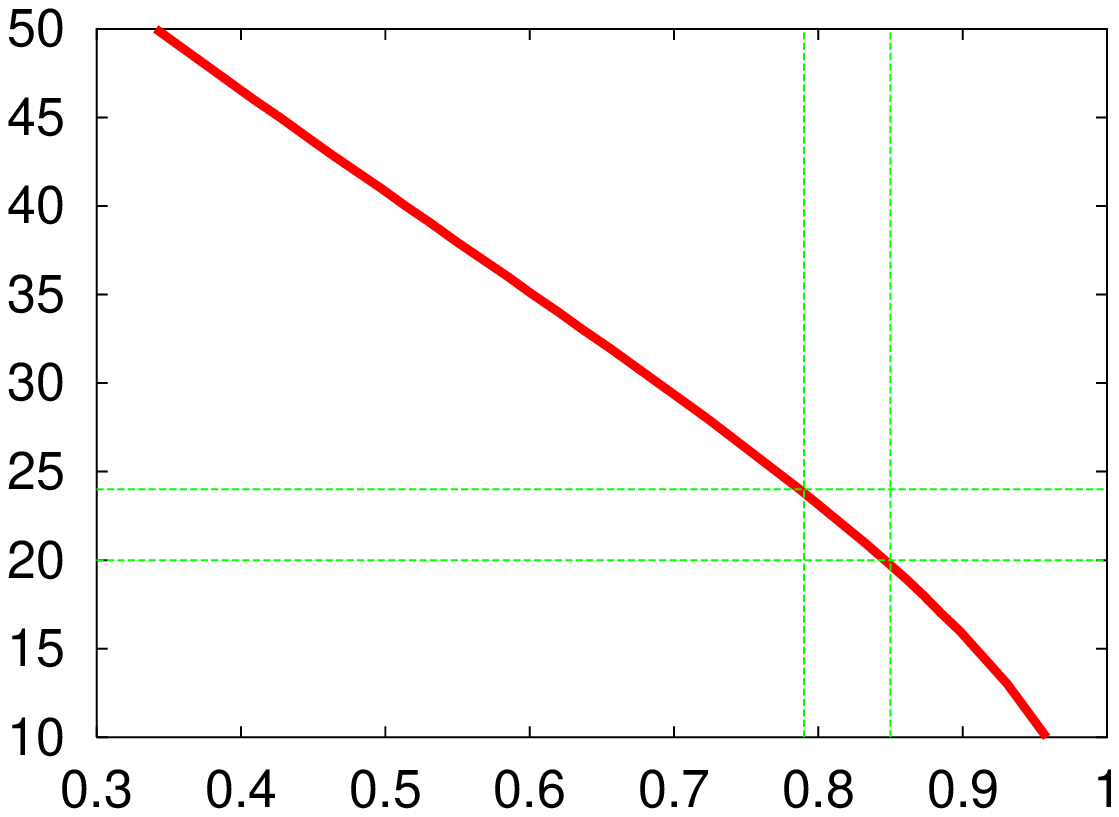}}}
\end{picture}\vspace{3cm}
\caption{a) Determining the $\tau$ polarisation
$P_{\tilde{\tau}_1\to\tau}$ from the pion energy distribution in the decay
$\tilde{\tau}_1\to \tau\tilde{\chi}^0_1\to \nu_{\tau} \pi
\tilde{\chi}^0_1$ leads to b)
an accurate determination of high $\tan\beta$: e.g. $\tan\beta=20\pm 2$
\cite{Stau}.}
\label{fig_tau}
\end{figure}

The polarisation of $\tau$'s from
$\tilde{\tau}_i\to\tau\tilde{\chi}^0_j$ is sensitive to $\tan\beta$ 
\cite{Noji}
and the $\tau$ polarisation can be rather accurately measured at a
LC via e.g. the $\tau$ decays into $\pi$'s, see Fig.~\ref{fig_tau}a, 
\cite{Stau}. 
It has been discussed that in 
case of a sufficient higgsino admixture in the
$\tilde{\chi}^0_j$ it is even possible to determine high $\tan\beta$
as well as $A_{\tau}$, without any assumptions on the SUSY breaking
mechanism: after determining the $\tilde{\tau}$ mixing
angle via polarised rates, preferable in the configuration
$\sigma_{RL}$ due to $WW$ background suppression,
one can determine $\tan\beta$ from the $\tau$
polarisation in the decay $\tilde{\tau}_1\to \tau \tilde{\chi}^0_1$, see
Fig.\ref{fig_tau}b.  Even for high $\tan\beta\ge 20$ one can reach
an accuracy of about $10\%$.

\subsubsection{Extended SUSY models}
In case of e.g. R--parity violating SUSY non--standard couplings could
occur which produce a scalar particle in the s--channel:
$e^+e^-\to\tilde{\nu}\to e^+e^-$. The process gives a significant signal
over the background. Since it requires both left--handed $e^-$ and
$e^+$ beams -- the $LL$ configuration-- it can be easily analysed and 
identified
by the use of beam polarisation
(\cite{Spiesberger,Steiner}): here 
simultaneously polarised beams
enhance the signal by about a factor of 10, see Table~\ref{tab_rp}.
\begin{table}
\begin{center}
\begin{tabular}{|l||c|c|}
\hline
 & $\sigma(e^+e^-\to e^+ e^-)$ with &  Bhabha--background\\
& $\sigma(e^+ e^-\to \tilde{\nu}\to e^+ e^-)$&  \\ \hline
unpolarised & 7.17 pb & 4.50 pb \\
$P_{e^-}=-80\%$ & 7.32 pb & 4.63 pb\\
$P_{e^-}=-80\%$, $P_{e^+}=-60\%$ & 8.66 pb & 4.69 pb\\
$P_{e^-}=-80\%$, $P_{e^+}=+60\%$ & 5.97 pb  & 4.58 pb\\
\hline
\end{tabular}
\caption{Sneutrino production in
R--parity violating SUSY:
Cross sections of $e^+ e^-\to \tilde{\nu}\to e^+ e^-$ for
  unpolarised beams, $P_{e^-}=-80\%$ and unpolarised positrons and
$P_{e^-}=-80\%$, $P_{e^+}=-60\%$. The study was made for
$m_{\tilde{\nu}}=650$~GeV, $\Gamma_{\tilde{\nu}}=1$~GeV, an angle cut
of $45^0\le \theta \le 135^0$ and the R--parity violating coupling
$\lambda_{131}=0.05$ \cite{Spiesberger}.\label{tab_rp}}
\end{center}
\end{table}

\begin{figure}[htb]
\begin{picture}(12,100)
\setlength{\unitlength}{1cm}
\put(0,-1){\mbox{
\includegraphics[height=.21\textheight,width=.33\textheight]{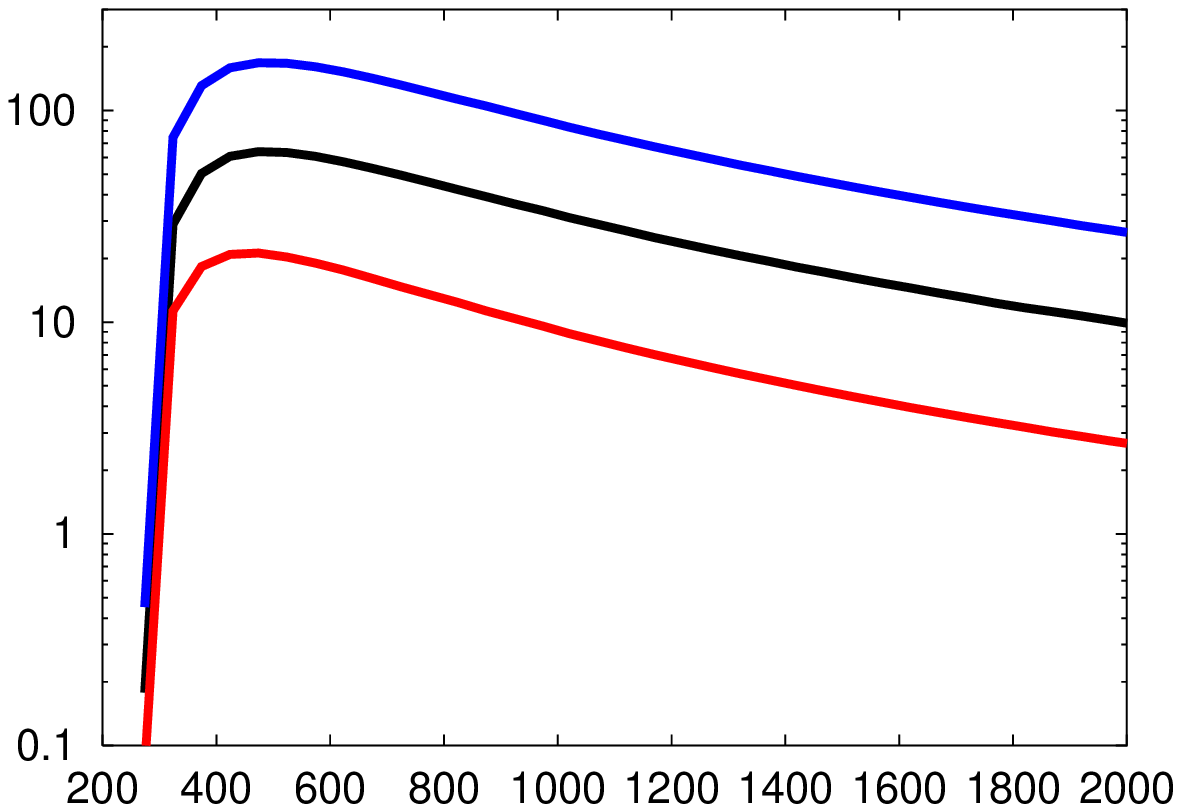}}}
\put(4.8,3){\small MSSM}
\put(4,2.2){\small $(-+)$}
\put(4.1,1.2){\small $(00)$}
\put(4,.5){\small $(+-)$}
\put(0,3){\small $\sigma(\tilde{\chi}^0_1\tilde{\chi}^0_2)$/fb}
\put(5,-1.3){\small $\sqrt{s}$~GeV}
\put(6.7,3){\small $\sigma(\tilde{\chi}^0_1\tilde{\chi}^0_2)$/fb}
\put(11.3,-1.3){\small $\sqrt{s}$~GeV}
\put(10.1,2.3){\small $(+-)$}
\put(10.2,1.1){\small $(00)$}
\put(10.1,0){\small $(-+)$}
\put(10.6,3){\small  (M+1)SSM}
\put(3.3,3){\small a)}
\put(9.5,3){\small b)}
\put(6.5,-.9){\mbox{\includegraphics[height=.19\textheight,
width=.32\textheight]{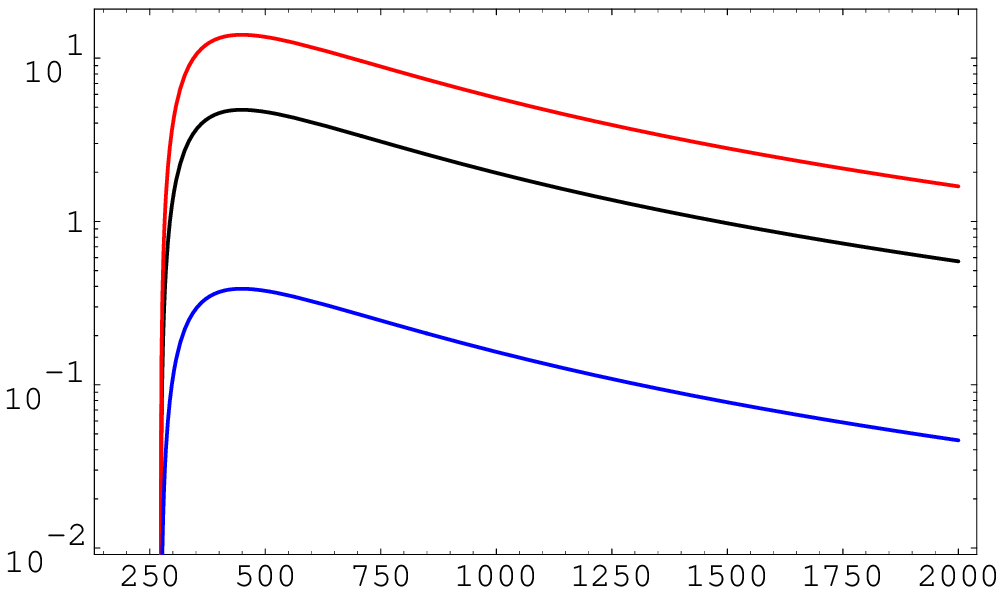}}}
\end{picture}\vspace{1.3cm}
\caption{Cross sections for the process $\sigma(e^+e^-\to
\tilde{\chi}^0_1\tilde{\chi}^0_2)$ with polarised beams $(P_{e^-}=\pm 80\%,
P_{e^+}=\mp 60\%)$ for an example in a) the MSSM and in b) the
(M+1)SSM, where the mass spectra of the light neutralinos are
similar \cite{Hesselbach}. \label{fig_ex}}
\end{figure}

One could also extend the MSSM without changing the gauge group, by
introducing an additional Higgs singlet:
it leads to the (M+1)SSM with one additional neutralino. 
Since the mass spectra of
the four light neutralinos could be similar to those in the MSSM in 
some parts of the (M+1)SSM/MSSM parameter space, a
distinction between these models might be difficult via spectra and 
rates alone.
However, polarisation effects might then
indicate the different coupling structure in the (M+1)SSM \cite{Hesselbach}
and help disentangling the models, Fig.~\ref{fig_ex}.

\subsection{Large extra dimensions}
Another approach for physics beyond the SM, which 
could also resolve the hierachy problem, is the introduction
of large extra dimensions. At a LC the process $e^+e^-\to \gamma G$ 
is promising and it has been worked out that with running on two
different $\sqrt{s}$ one can determine the number of extra dimensions
\cite{TDR,Wilson}. The use of polarised beams in this context enlarges on
one hand the sensitivity to the new scale $M_{*}$ and suppresses
on the other hand
the main background $e^+e^-\to \nu \nu \gamma$ significantly. 
The ratio $S/\sqrt{B}$ is enhanced by a factor of about 2.1 (4.4) if
$P_{e^-}=+80\%$ (and $P_{e^+}=-60\%$) is used.

\section{Summary}
A Linear Collider in the TeV range with its 
clean initial state of $e^+ e^-$ collisions is 
ideally suited for the search for new physics, for the determination of 
both 
Standard Model and non-standard couplings with high precision and for 
revealing 
the structure of the underlying model.
The use of polarised beams plays a decisive role in this context.
We have shown that simultaneous polarisation of both beams 
can significantly expand the accessible physics 
opportunities compared to the case of $e^-$ polarisation 
only\footnote{For updates 
see POWER group (Polarisation at Work in Energetic Reactions)\\ 
\phantom{LHC}http://www.ippp.dur.ac.uk/ $\tilde{}$ gudrid/power/}.  
The use of simultaneously polarised
$e^-$, $e^+$ beams has several advantages for: 
determining quantum numbers of new particles,
providing higher sensitivity to
non--standard couplings,
increasing rates and background suppression, 
raising the effective luminosity
and expanding the range of measurable experimental observables e.g.
with the help of transversely polarised beams. 

\vspace{.5cm}
The author would like to thank Yousef Makdisi with his nice and friendly
organising team for a wonderful and very interesting conference! 
G.M.--P. was partially supported by BNL.


\end{document}